\documentclass[printer]{aa}
\usepackage{natbib}
\usepackage{bm}
\usepackage{amssymb}
\usepackage{graphicx}
\usepackage{hyperref}
\usepackage{comment}
\usepackage[caption=false]{subfig}
\graphicspath{{./figures/}}

\newcommand{\rhoism}{\mbox{$\rho_{\rm ISM} $}}
\newcommand{\vism}{\mbox{$v_{\rm ISM}$}}
\newcommand{\vorb}{\mbox{$v_{\rm orb}$}}

\newcommand{\Msun}{\mbox{$\mathrm{M}_{\odot}$}}

\makeatletter
\newcommand\footnoteref[1]{\protected@xdef\@thefnmark{\ref{#1}}\@footnotemark}
\makeatother

\begin{document}

\title{Changes in orientation and shape of protoplanetary discs moving through an ambient medium}
\titlerunning{Orientation and shape of protoplanetary discs}
\author{T.P.G. Wijnen\inst{1,2}, F.I. Pelupessy\inst{2,3}, O.R. Pols\inst{1}, S. Portegies Zwart\inst{2}}
\authorrunning{T.P.G. Wijnen et al.}
\institute{Department of Astrophysics/IMAPP, Radboud University Nijmegen, P.O. Box 9010, 6500 GL Nijmegen, The Netherlands\\
\email{thomas.wijnen@astro.ru.nl}
\and Leiden Observatory, Leiden University, PO Box 9513, 2300 RA Leiden, The Netherlands
\and CWI, P.O. Box 94079, 1090 GB Amsterdam, The Netherlands
\offprints{T.P.G. Wijnen}
}
\date{Received ..../ Accepted ....}

\abstract{Misalignments between the orbital planes of planets and the equatorial planes of their host stars have been observed in our solar system, in transiting exoplanets, and in the orbital planes of debris discs. We present a mechanism that causes such a spin-orbit misalignment for a protoplanetary disc due to its movement through an ambient medium. Our physical explanation of the mechanism is based on the theoretical solutions to the Stark problem. We test this idea by performing self-consistent hydrodynamical simulations and simplified gravitational $N$-body simulations. The $N$-body model reduces the mechanism to the relevant physical processes. The hydrodynamical simulations show the mechanism in its full extent, including gas-dynamical and viscous processes in the disc which are not included in the theoretical framework. We find that a protoplanetary disc embedded in a flow changes its orientation as its angular momentum vector tends to align parallel to the relative velocity vector. Due to the force exerted by the flow, orbits in the disc become eccentric, which produces a net torque and consequentially changes the orbital inclination. The tilting of the disc causes it to contract. Apart from becoming lopsided, the gaseous disc also forms a spiral arm even if the inclination does not change substantially. The process is most effective at high velocities and observational signatures are therefore mostly expected in massive star-forming regions and around winds or supernova ejecta. Our $N$-body model indicates that the interaction with supernova ejecta is a viable explanation for the observed spin-orbit misalignment in our solar system.}

\maketitle

\section{Introduction}

Intuitively, one might expect that the spin axis of a star should be  aligned with the orbital axis of its protoplanetary disc and the planets that form there. However, the solar system is known to exhibit a misalignment of roughly 7$^{\circ}$ between the equatorial plane of the sun and the orbital planes of its planets \citep{beck05}. Such spin-orbit misalignments have also been observed for transiting exoplanets \citep{hebrard08, johnson09, gillon09, narita09, pont09, winn09} and for debris discs \citep{watson11, greaves14}. For debris discs, i.e. protoplanetary discs cleared of the majority of their gas, the observed misalignment is not greater than 30$^{\circ}$. Several scenarios have been put forward to explain these misalignments, ranging from interaction with stellar and/or planetary companions \citep[see e.g.][]{fabrycky07, nagasawa08, matsakos16, hamers17} to magnetic interactions between the star and its disc \citep{lai11} and to primordial misalignment \citep{bate10, fielding15}.

In addition, transition discs, i.e. protoplanetary discs with inner clearings, are known to show asymmetries \citep[e.g.][]{oppenheimer08, isella12, van_der_marel13} and spiral structures \citep[e.g.][]{pietu05, corder05, muto12, christiaens14, van_der_marel16}. These features are generally explained by the presence of a planet that can cause the Rossby wave instability \citep{lovelace99, de_val-borro07} or trigger density waves \citep{kley12}. Although the origins of these observed structures in transition discs and the spin-orbit misalignment in debris discs and planetary systems are generally investigated separately, we will show in this work that they could be related. As a star surrounded by a protoplanetary disc moves through an ambient medium, the experienced drag and interaction with the interstellar medium (ISM) can affect the orientation and evolution of the protoplanetary disc. In previous works \citep[][hereafter paper I and II]{wijnen16, wijnen17}, we found that the accretion of gas from the ISM with no azimuthal angular momentum causes the disc to contract because the specific angular momentum of the disc decreases. Interactions of debris discs with their environments have also been observed. For example, \citet{maness09} and \citet{debes09} respectively find that the morphology and asymmetry observed in the dusty debris disc HD 61005 and the asymmetries and warping in the debris disc HD 32297 can be explained via interaction with the ambient interstellar medium. Our solar system is believed to have experienced interactions with either the wind of an evolved asymptotic giant branch star \citep[e.g.][]{busso99} or, more likely, supernova ejecta \citep[e.g.][]{clayton77}. The observed misalignment in our solar system may be related to these interactions. 

The interaction of dust grains in a debris disc with the ISM has been investigated in terms of the Stark problem (also known as the accelerated Kepler problem, see e.g. \citealt{belyaev10, lantoine11, pastor12} and Sect. \ref{sec:stark_tilt}), sandblasting of the disc by the ISM \citep{artymowicz97}, and with $N$-body integrations \citep{debes09, marzari11}. For example, \citet{marzari11} find that asymmetries in debris discs can be attributed to interaction with an ISM flux if the optical depth of the disc is low, i.e. the collisional lifetimes of dust particles in the disc are long. Several authors \citep[see e.g.][]{namouni05, namouni07, pastor12} have derived secular time derivatives of Keplerian orbital elements of spherical dust particles subject to a small constant mono-directional force caused by an ISM flow. Here we investigate star-disc alignment as a result of interaction with an ambient medium by performing self-consistent hydrodynamical simulations of an inclined gaseous disc subject to an incoming gas stream. 

In this paper we show that, in addition to affecting the outer regions and size of a protoplanetary disc, an ISM flow may induce a tilt of the disc that is perpendicular to the flow direction. This process occurs even if the geometry of the disc is initially symmetric. It may not be immediately apparent why this happens since no net torque should be present on a symmetric disc. We show that a tilt is expected from the effects of the ISM force on the orbits of the gas particles within the disc and can be qualitatively described by the solutions to the Stark problem. We perform hydrodynamic as well as gravitational $N$-body simulations, and compare their outcome with a theoretical description of the process. In addition, we derive the time scale on which the disc changes its orientation as a function of the ISM density and relative velocity, and investigate under which conditions this mechanism affects the orientation of the disc. We show that this process can be linked to the formation of asymmetries and spiral structures in the disc. We discuss the relevant theoretical processes in Sect. \ref{sec:theory_tilt}. The set-up of our simulations is presented in Sect. \ref{sec:setup_tilt} and the results in Sect. \ref{sec:results_tilt}. In Sect. \ref{sec:discussion_tilt} we discuss our assumptions and relate the implications of the mechanism to observations and the required physical conditions.

\section{Theory}\label{sec:theory_tilt}

A qualitative theoretical description of the physical mechanism that changes the orientation of the disc can be provided in the context of the Stark problem. The Stark problem has the same physical basis as the Stark effect where the shifting and splitting of spectral lines is caused by an external electric field \citep{stark14}. We use the theoretical solutions to this problem to explain the behaviour of the disc in the hydrodynamical simulations.

\subsection{Stark problem}\label{sec:stark_tilt}

\begin{figure}
\centering
    \includegraphics[width=0.49\textwidth]{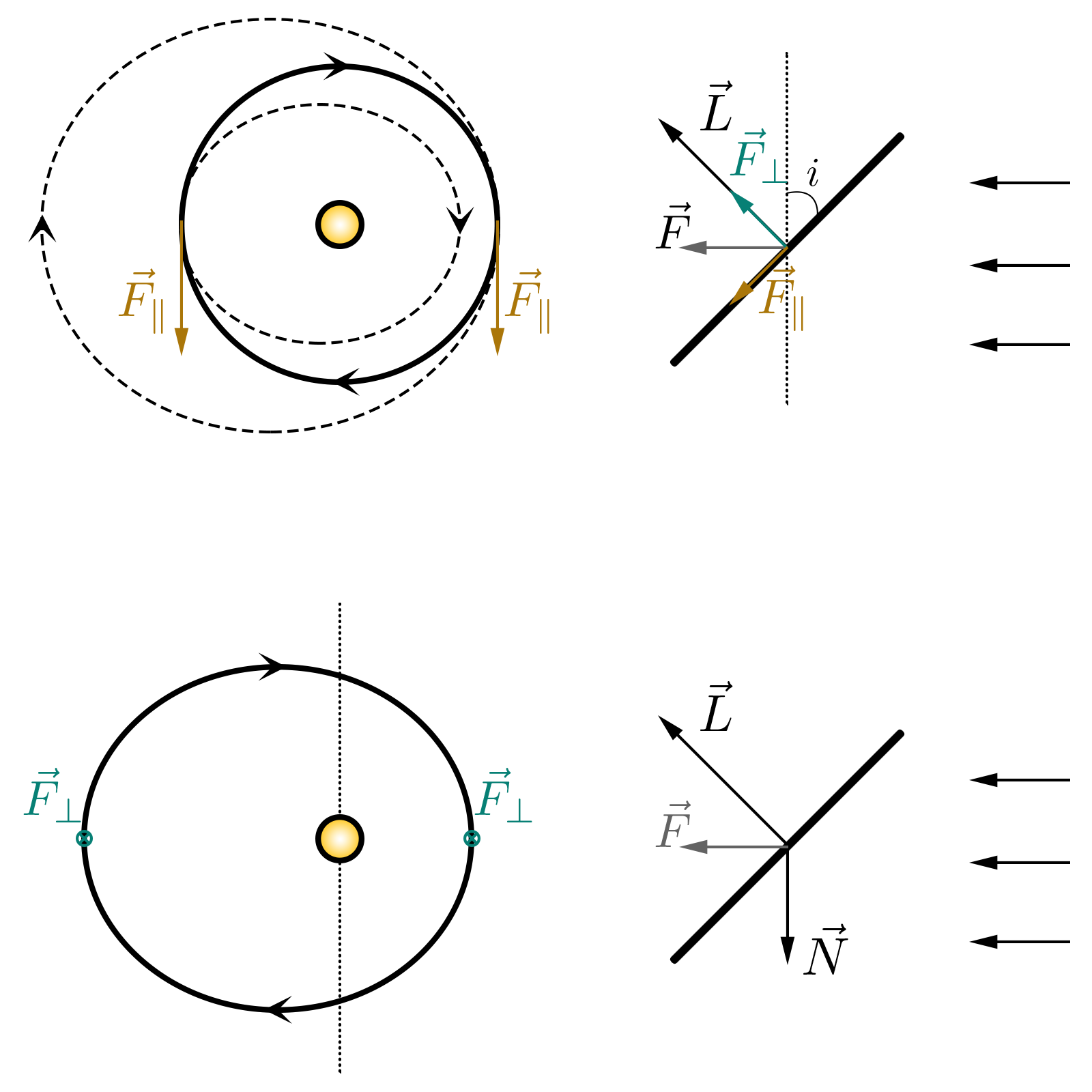}
    \caption{Schematic overview of the mechanism that causes the net torque on the disc. The figure is not to scale. The left side shows the orbit of a particle in the plane of the disc and the right side shows the edge-on view of the disc. In this overview the ISM flows in from the right.  See Sect. \ref{sec:stark_tilt} for an explanation of the mechanism that tilts the disc. \label{fig:stark_tilt}}  
\end{figure}

The Stark problem is a classical two-body problem in which a particle moves in a potential, for example the gravitational potential of a star, and is subject to a constant mono-directional force. In the context of our work this force is exerted by an ISM flow. In Fig. \ref{fig:stark_tilt} we have drawn a schematic overview of the behaviour of particles when they are subject to a force exerted by the ISM flow. We can qualitatively understand the process by considering a disc consisting of particles that initially move on circular orbits. The particles represent a certain mass and volume and the normal of their orbital plane is inclined by an angle $i_0$ with respect to the force vector. The disc rotates clockwise and we have indicated its angular momentum vector with $\overrightarrow{L}$. The left half of Fig. \ref{fig:stark_tilt} shows the orbit of a particle in the plane of the disc, i.e. viewed along the orbital axis of the disc (in the direction of the angular momentum vector $\overrightarrow{L}$). The right half of the figure shows the edge-on view of the orbit. In this view, the left side of the orbit is the near side and comes out of the plane of the paper. The ISM flow exerts a force everywhere on the particles that can be split into a component parallel, $\overrightarrow{F_{\parallel}}$, and a component perpendicular, $\overrightarrow{F_{\bot}}$, to the plane of the orbit. When the orbital velocity of the particle has a component parallel to $\overrightarrow{F_{\parallel}}$, the velocity of the particle increases and so will the eccentricity and  apocentre of its orbit. Likewise, if the orbital velocity has a component anti-parallel to $\overrightarrow{F_{\parallel}}$ the velocity of the particle is reduced, thereby increasing the eccentricity and decreasing its pericentre distance. This is shown schematically in the top left panel of Fig.~ \ref{fig:stark_tilt} for the two points in the orbit where the effect is strongest. This process will result in an eccentric orbit, i.e. a lopsided disc where, in this example, the particle spends a larger fraction of the orbital period on the left side of the star than on the right side (see lower left of the figure). As a result, the force exerted on a particle while it is on the left side of the orbit has a longer arm with respect to the star and acts for a longer time than the force exerted during the transit of the right side. This produces a net torque, $\overrightarrow{N}$, on the time-averaged orbit of the particle which is orthogonal to the direction of the force. This is depicted in the bottom right. Since $\overrightarrow{N}=d\overrightarrow{L}/dt$, the orbital plane is tilted in a direction perpendicular to the force vector. In the absence of other forces the eccentricity will keep growing to a maximum value when the orbital plane has reached perpendicular alignment to the force vector. The torque is thus maintained and the orbit will continue to tilt in the same direction.
Now, $\overrightarrow{F_{\parallel}}$ points in the opposite direction with respect to the orientation of the orbit and the eccentricity will decrease. At the moment the eccentricity becomes zero, the inclination attains a maximum. The configuration of the disc now mirrors the initial situation and the process is reversed. The force increases the eccentricity again, but this time the apocentre of the orbit will be on the right side of the star, i.e. the argument of pericentre, $\omega$, has changed by $180^\circ$. Therefore, $\overrightarrow{N}$ points upwards and once more the orbital plane will be tilted perpendicular to the force vector, thereby decreasing the inclination. Both the eccentricity and inclination of the orbit will oscillate back and forth between a minimum and maximum. We  refer to this process as the tilt process. 

In the case of the classical Stark problem,the semi-major axis $a$ of the orbit remains constant and the oscillation of the eccentricity and inclination can be described analytically as a function of time. For the case where the initial orbit is circular and the normal of the orbital plane is inclined by an angle $i_0$ with respect to the force, the eccentricity can be derived from (cf. \citealt{namouni05, namouni07})
\begin{equation}\label{eq:dedt_tilt}
\frac{de}{dt} =\frac{3}{2} \sqrt{\frac{a}{GM_*}} A(t) \epsilon\sqrt{\sin{i_0}^2 -e^2}, 
\end{equation}
where $G$ is the gravitational constant, $M_{*}$ the mass of the star, $\epsilon$ is the sign of $\cos \omega$, and $A(t)$ the acceleration, which we allow to vary in time (it is constant in the classical Stark problem). Integrating Eq. \ref{eq:dedt_tilt} yields
\begin{equation}\label{eq:ecc_tilt}
e(t) =\left| \sin \left(\frac{3}{2} \sqrt{\frac{a}{GM_*}} \int_0^t A(t') dt' \right) \sin{i_0} \right|.
\end{equation}
The inclination $i$ can be obtained via
\begin{equation}\label{eq:inc_tilt}
\cos i(t) = \frac{\cos i_0}{\sqrt{1-e(t)^2}},
\end{equation}
from the conservation of the angular momentum component along the direction of the force. 

In the solutions to the Stark problem given above, orbit averaging is applied. The description therefore also applies when considering the rings of a disc that move on Kepler orbits.

\subsection{Tilt process for a gaseous disc}\label{sec:disctheory_tilt}

In the case of a gaseous disc embedded in an ISM flow, the force is exerted by the ram pressure of the flow. In contrast to the classical Stark problem, $\overrightarrow{F_{\parallel}}$ is not constant but depends on the relative velocity of the gas particles in the disc with respect to the velocity of the flow. As a consequence, the acceleration on a gas particle in the disc depends on its location in the disc and the component of its orbital velocity parallel to the ISM velocity, $v_{\parallel}$. Qualitatively, however, the effect of the force exerted by the flow is the same as described above: the decrease in the force when $v_{\parallel}$ is in the same direction as $\vism$ is compensated by an increase in the force when $v_{\parallel}$ opposes $\vism$. For the present purposes, we can assume that the disc consists of an incompressible gas. A complete derivation would require solving the fluid equations for conservation of angular momentum \citep{olbers12}, which is non-trivial for a disc that becomes eccentric. The acceleration can then be written as
\begin{equation}\label{eq:at_sph_general_tilt}
A(r,\phi, t) = \frac{\rhoism \left(\vism - v_{\parallel}(r,\phi)\right)^2 \cos i(t)}{\Sigma(r, \phi, t)},
\end{equation}
where $r$ and $\phi$ are cylindrical coordinates, $\rhoism$ and $\vism$ are the density and velocity of the ISM flow, $\Sigma(r, \phi, t)$ is the surface density profile of the disc, and the factor $\cos i(t)$ arises from accounting for the inclined surface area. 

Apart from the loss of axisymmetry due to the non-constant force, the description in Sect.~\ref{sec:stark_tilt} is complicated by two additional effects: (1) the accretion of ISM causes the disc to contract so that the orbital semi-major axes of the particles are no longer constant and (2) gas particles in the disc do not orbit independently from each other because the gas interacts gravitationally, hydro-dynamically, and viscously. Considering the first effect, we  showed in paper II that in case of axisymmetry $da/dt$ depends on the change in the surface density profile of the disc and that there is no analytic solution to describe $a$ as a function of time. Furthermore, $A(r, \phi, t)$ depends on the eccentricity via $\cos i(t)$ and Eq. \ref{eq:inc_tilt}. Since, via the surface density profile, $a(t)$ also depends on the eccentricity, integrating Eq. \ref{eq:dedt_tilt} no longer results in Eq. \ref{eq:ecc_tilt} and the coupled differential equations for $da/dt$, $de/dt$, and $d\omega/dt$ \citep[see e.g.][]{namouni05} would have to be solved simultaneously. Second, gas dynamical and viscous interactions in the disc have not been taken into account. It may be sufficient to describe $de/dt$ in a gaseous disc by including a dissipation term in Eq. \ref{eq:dedt_tilt} to account for these interactions. The dissipation will damp the amplitude of the oscillation in the eccentricity and inclination. Unfortunately, little is known about this dissipation and hence we  cannot include a dissipation term in Eq.~\ref{eq:dedt_tilt}. Considering this and the above-mentioned point that the assumption of axisymmetry is no longer valid, solving the coupled differential equations would not yield a comprehensive physical description of the tilt process in a gaseous disc. 

In order to obtain a qualitative description of the process to which we can compare our hydrodynamical simulations, we use several simplifying assumptions. By neglecting dissipation and the component of the force that depends on the location in the disc, assuming axisymmetry and treating $\cos i$ as constant in time, we can write
\begin{equation}\label{eq:at_sph_tilt}
A(r,t) = \frac{\rhoism \vism^2 \cos i_0}{\Sigma(r,t)}.
\end{equation}
Using the theoretical model for ISM accretion derived in paper II and assuming that $\rhoism$ and  $\vism$ are constant, we can write the surface density profile as
\begin{equation}\label{eq:surfdens_tilt}
\Sigma(r,t) = \Sigma_0 (r/r_0)^{-n} + 5 \rhoism \vism \cos{i_0} \, t
\end{equation}
where we  assume that the surface density profile at $t=0$ can be written as $\Sigma_0 (r/r_0)^{-n}$ and we take the perpendicular component of $\vism$ as discussed in Sect. 5.5 of paper II. With the simplifications above we can write
\begin{equation}\label{eq:acc_sph_tilt}
\int_0^t A(r, t') dt' = \frac{1}{5}\vism \ln \left| \frac{\Sigma(r,t)}{\Sigma(r,t=0)} \right|,
\end{equation}
and we use Eqs.~\ref{eq:ecc_tilt} and \ref{eq:inc_tilt} to qualitatively describe the behaviour of the disc. Although physically this is not correct, we show in the results section that this still provides an insightful description of the physical process.

We estimate a time scale for the change in eccentricity, $\tau_e$, at $t=0$ from Eq.~\ref{eq:dedt_tilt}, assuming that $\tau_e = (de/dt)^{-1}$ and that the disc is axisymmetric. Using Eq.~\ref{eq:at_sph_tilt} this yields
\begin{equation}\label{eq:taue_tilt}
\tau_e (r) = \frac{2 \vorb(r) \Sigma(r)}{3 \rhoism \vism^2 \cos{i_0} \sin{i_0}},
\end{equation}
with $\vorb(r) = \sqrt{G M_*/r}$. Since the changes in inclination and eccentricity are coupled, the inclination varies on the same time scale. We can relate $\tau_e$ to the time scale of mass loading onto the disc, i.e. Eq. 8 from paper II
\begin{equation}\label{eq:taum_tilt}
\tau_{\dot{m}} (r) = \frac{\Sigma(r)}{5\rhoism \vism \cos{i_0}},
\end{equation}
to derive for which $\vorb(r)$, i.e. at which $r$, the eccentricity changes faster than the disc contracts. Equating Eqs.~\ref{eq:taue_tilt} and \ref{eq:taum_tilt} gives $\vorb(r) < 3/10\,\vism \sin{i_0}$. For a given momentum flux from the ISM, the tilting is thus more effective at high incoming velocity rather than at high ambient density. This is a consequence of the dependence of the process on the ram pressure, $\rhoism \vism^2$, so that increasing $\vism$ has a stronger effect than increasing $\rhoism$. 

The mechanism described above implies that the tilting of the disc happens without any precession and that the component of the angular momentum parallel to the velocity and force vector remains unaffected by this process. In the next section we  verify the qualitative theoretical predictions of this mechanism using numerical simulations.

\section{Numerical set-up}\label{sec:setup_tilt}

\begin{table*}
\begin{center}
\caption{Initial conditions for our simulations. Only the density and velocity vary between simulations, which are listed in Table \ref{tb:models_tilt}.}
\label{tb:parameters_tilt}
\begin{tabular}{lrl}
\hline
\textbf{Parameter}&\textbf{Value}&\textbf{Description}\\
\hline
$\mathrm{N_{\rm disc}}$& 128 000&Number of disc particles\\
$\mu$&2.3&Mean molecular weight\\
$M_*$&$0.4\,\Msun$&\\
$f_{\rm disc}$&$0.01$&$\frac{M_{\rm disc}}{M_*}$\\
%$M_{\rm disc}$&$0.004\,\Msun$&\\
$R_{\rm disc, inner}$&10\,AU&\\
$R_{\rm disc, outer}$&100\,AU&\\
$\Sigma(r)$&$\Sigma_0(\frac{r}{r_0})^{-1.5}$&Surface density profile\\
EoS&Isothermal&Equation of state\\
$T$&25\,K&Temperature of gas particles\\
$c_s$&$0.3\,\mathrm{km\, s^{-1}}$&Sound speed\\
$R_{\rm sink}$&1\,AU&Sink particle radius\\
$N_{\rm neighbours}$& $64\pm2$&\\
$\epsilon_{\rm grav}$&$1\,\mathrm{AU}$&Gravitational softening length\\
$\alpha_{\rm SPH}$&0.1&Artificial viscosity parameter\\
$\beta_{\rm SPH}$&1&Artificial viscosity parameter\\
$L_{\rm cylinder}$& 400 AU&Length of computational domain\\
$R_{\rm cylinder}$& 200 AU&Radius of computational domain\\
$t_{\rm sim}$&10 000 yr&Duration of the simulation\\
\hline
\end{tabular}
\medskip
\end{center}
\end{table*}

We investigated the influence of an ISM flow on the disc by performing hydrodynamical simulations of a self-gravitating gas disc embedded in an ISM flow. We compared these simulations to an $N$-body model in which we approximated the dynamics by considering the orbits of test particles (representing the gas disc) subjected to the gravity of the central star and the momentum and mass flux imparted by the ISM flow. Thus, in the hydrodynamical simulations, the tilt process occurs self-consistently as a result of interaction with the inflowing ISM, while in the $N$-body simulations the tilt process is induced by applying an artificial pressure force on the particles.

\subsection{Hydrodynamical simulations}\label{sec:setup_sph_tilt}
We performed smoothed particle hydrodynamic (SPH) simulations for which we used the same set-up as outlined in paper II. Our computational domain spans a cylinder with a length of 400 AU and a radius of 200 AU. The disc, consisting of 128 000 SPH particles, is positioned in the middle of the cylinder. A schematic overview of the set-up is shown in Fig. 1 in paper I. In contrast to papers I and II, we gave the disc an inclination $i$ of 45$^\circ$ with respect to the radial axis (see  top right panel of Fig. \ref{fig:stark_tilt}). As described in paper I, we constructed the incoming gas stream by adding a slice of new particles at the inflow boundary of the domain every time step. We used the SPH code Fi \citep{pelupessy04} within the AMUSE framework \citep{portegies_zwart13, pelupessy13}\footnote{\url{http://www.amusecode.org}}. For a detailed description of the set-up we refer to Sect. 3 of paper I and Sect. 3 of paper II. We used the same initial conditions and parameters as in paper II, which are repeated here in Table \ref{tb:parameters_tilt}. Fi uses the \citet{monaghan83} prescription for the viscosity factor $\Pi_{ij}$, but takes the minimum of the density and the maximum of the sound speed between two neighbours instead of the mean. Therefore, Fi is able to resolve shocks even at a relatively low value of the artificial viscosity parameter $\alpha_{\rm SPH}$. Our assumed value of $\alpha_{\rm SPH} = 0.1$ corresponds to a physical $\alpha$ of roughly 0.02 in the \citet{shakura73} prescription \citep[following][]{artymowicz94}. To distinguish the disc from the flow we used the clump finding algorithm Hop \citep{eisenstein98} in the parameterspace of $v_{\theta}$, $\rho$, and $v_x$ (see Sect. 3 of paper II).  For the SPH particles that are assigned to the disc by this algorithm and are gravitationally bound to the star, we calculated the orbital eccentricity and inclination from their orbital energy and angular momentum assuming they follow Kepler orbits. The orbital parameters are used in the analysis in Sect. \ref{sec:results_tilt}. We defined the overall eccentricity and inclination of the disc or a ring within the disc at any moment in time as the average of that parameter over the respective particles at that moment.

\subsection{$N$-body model}\label{sec:setup_nbody_tilt}
We used the code Huayno \citep{pelupessy12} within the AMUSE framework to perform the $N$-body simulations. Here we represent the disc by test particles that feel the gravity of the central star and have their velocity and mass $m$ changed by the captured mass and momentum from the ISM flow. We changed the mass of the $N$-body particles to account for the accretion that occurs in the gaseous disc, as described in paper II. The particles accrete mass and momentum according to assigned cross sections, assuming they are distributed in a flat disc. In the $N$-body simulations, the disc consists of 5000 particles that are distributed according to the parameters listed in Table \ref{tb:parameters_tilt}. We assigned each particle a fixed cross section $\sigma$ depending on its radius, using $\sigma(r) = \sigma_{\rm disc} (r/R_{\rm disc, inner})^{-1.5} /\sum\nolimits_{p} (r/R_{\rm disc, inner})^{-1.5}$, where $\sigma_{\rm disc}$ is the total surface area of the disc and $\sum\nolimits_{p}$ is the sum over all particles, such that the sum of the individual cross sections equals the surface area of the disc. As in the SPH simulations, we started the simulations with the disc inclined by an angle $i=45^\circ$ with respect to the incoming flow. To mimic an ISM flow in the $N$-body simulation, directed along the x-axis, we increased the mass and velocity of each particle every time step using
\begin{equation}\label{eq:nbody-impuls_tilt}
\begin{aligned}
&\Delta m = \rhoism (\vism - v_x) \sigma \cos{i_0} \Delta t, \\ 
&\Delta v_x = \frac{\rhoism (\vism - v_x)^2 \sigma \cos{i_0} \Delta t}{m + \Delta m} &%\Delta p_x = \vism \Delta m
\end{aligned}
\end{equation}
for the particles with $v_x < \vism$. Multiplying by $\cos{i_0}$ instead of $\cos{i(t)}$ introduces a minor error compared to the one introduced by the fixed cross sections in time. Assigning new, physically consistent cross sections each time step is non-trivial as the surface density profile of the disc changes during the simulation, and we therefore decided to keep them constant. We experimented with cross sections that vary in time, but they provided practically the same outcome of the simulations. We calculated the eccentricity and inclination of the particles in the same way as we did for the SPH simulations. 

To provide a qualitative theoretical description of our $N$-body models, we assume that we can write the acceleration of a particle as
\begin{equation}\label{eq:acc_grav_tilt}
A(r, t)  = \frac{\rhoism \vism^2 \sigma(r) \cos{i_0}}{m(r, t)},
\end{equation}
analogously to Eq.~\ref{eq:at_sph_tilt}, but replacing $\Sigma(r,t)$ by $m(r, t)/\sigma(r)$, where $\sigma(r)$ is the cross section of the particle and its mass is given by $m(r, t)=m_0 + \rhoism \vism \sigma(r) \cos{i_0} t$. As discussed in Sect. \ref{sec:disctheory_tilt}, this does not provide a physically correct description of the evolution of the particles at any moment in time. However, we  show that for the first oscillation of the eccentricity and inclination in our $N$-body models this approximation is still adequate.

\subsection{ISM densities and velocities}

We chose six combinations of densities and velocities that overlap with those used in paper II. The lowest velocities that we used in paper II, 1 and 3 km/s, are in the regime where gravitational focusing, i.e. Bondi-Hoyle accretion, of ISM from radii $>R_{\rm disc}$ complicates the analysis. Furthermore, the change in eccentricity and inclination of the disc is expected to be stronger at a higher momentum flux. As a representative sample of velocities, we therefore used $v = 5$, 10, and 20 km/s. The simulations with the lowest number density we used in paper II, $n = 5 \times 10^4\,\mathrm{cm}^{-3}$, seemed to suffer from numerical issues (due to the low number of particles in the flow, see Sect. 5.1 of paper II).  We therefore decided to use $n = 5 \times 10^5$ and $5 \times 10^6\,\mathrm{cm}^{-3}$. For these number densities, $\ge 10^3\,\mathrm{cm}^{-3}$, and our adopted temperature of 25 K, the cooling time scale is $\lesssim 16$ years (see Sect. 3.1. of paper I). Assuming the same mean molecular weight of 2.3 as in paper I and II, these number densities correspond to mass densities of $1.9 \times 10^{-18}$ and $1.9 \times 10^{-17}$ g/cm$^3$. This results in six different models, which are listed in Table \ref{tb:models_tilt}. 

Our  aim was to understand the physical process that changes the orientation of the disc, as a function of the density and velocity of the ISM. We therefore started each simulation with an inclination of 45 degrees, which we regard as a representative value and which allowed us to disentangle the relevant physical mechanisms that govern the response of the disc to the ISM flow.

\begin{table}
\centering
\caption{Models used for our alignment study.}
\label{tb:models_tilt}
\begin{tabular}{lcccc}
\hline
\textbf{Label}&$\mathbf{v}$ [km/s]&$\mathbf{n}$ [$\mathrm{cm^{-3}}$]&$\bm{\rho}$ [$\mathrm{g/cm^3}$]&$\mathbf{i}$ [$^\circ$]\\
\hline
V5N5i45&5&$5 \times 10^5$&$1.9 \times 10^{-18}$&45\\
V5N6i45&5&$5 \times 10^6$&$1.9 \times 10^{-17}$&45\\
V10N5i45&10&$5 \times 10^5$&$1.9 \times 10^{-18}$&45\\
V10N6i45&10&$5 \times 10^6$&$1.9 \times 10^{-17}$&45\\
V20N5i45&20&$5 \times 10^5$&$1.9 \times 10^{-18}$&45\\
V20N6i45&20&$5 \times 10^6$&$1.9 \times 10^{-17}$&45\\
\hline
\end{tabular}\\
{\em Columns 1 to 5}:  The label we use to refer to the simulation, the velocity of the ISM, the number density of the ISM, the mass density of the ISM, and the inclination of the disc's axis with respect to the ISM flow. 
\end{table}

\section{Results}\label{sec:results_tilt}

We start by discussing simulation V20N6i45 (see Table \ref{tb:models_tilt}) by comparing the behaviour of different rings in the disc in the $N$-body and SPH simulations in Sect. \ref{sec:res_case_tilt}. Then we discuss the evolution of the disc as a whole in all simulations in Sect. \ref{sec:res_all_tilt} and we end with the evolution of the angular momentum vector in Sect. \ref{sec:res_am_tilt}.

\subsection{Evolution of concentric rings in the disc}\label{sec:res_case_tilt}

\begin{figure*}[!ht]
\centering
\includegraphics[width=\textwidth]{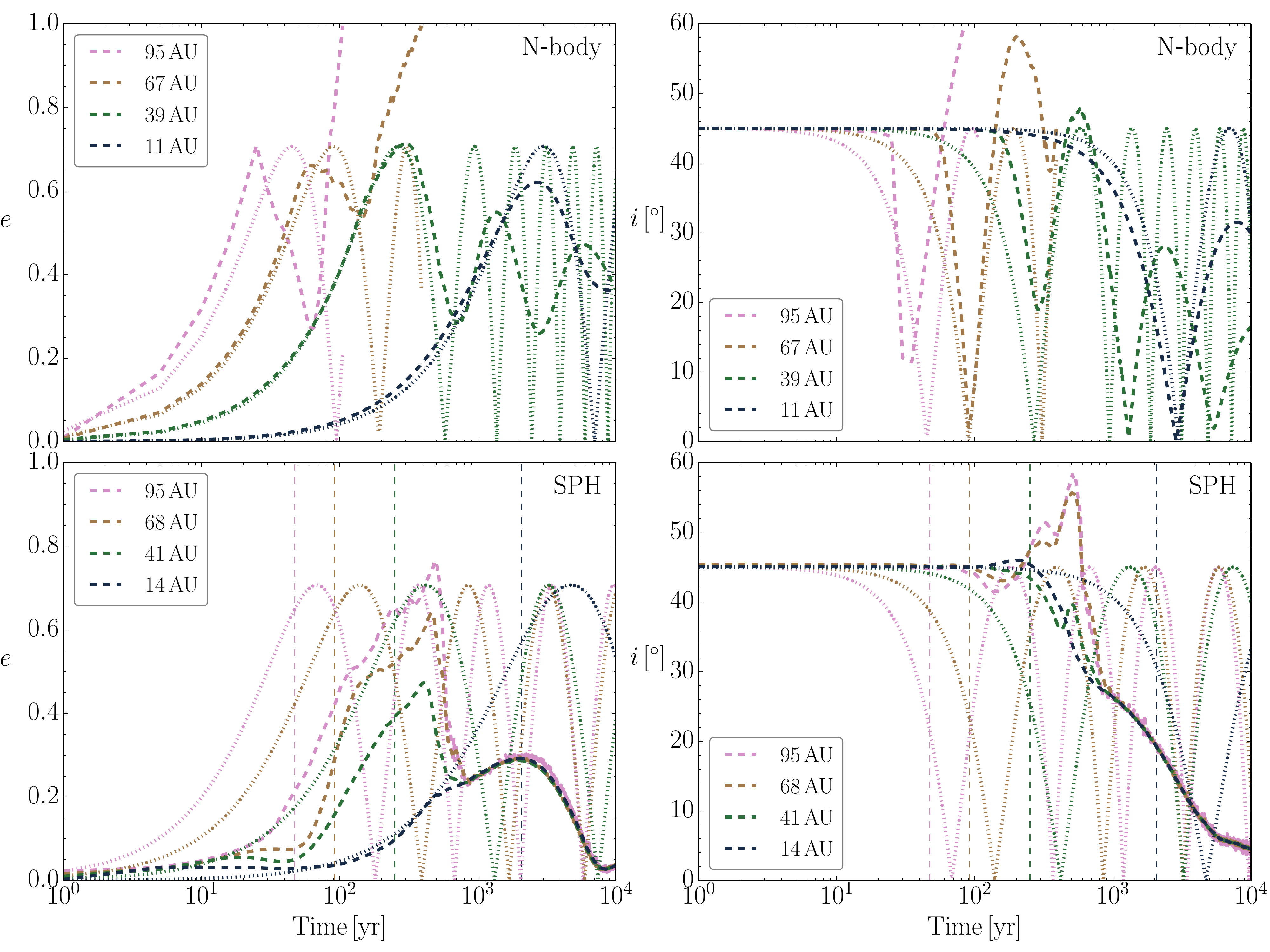}
\caption{At the start of the simulation, we binned the disc in both the $N$-body and SPH simulation V20N6i45 in ten concentric rings, based on the minimum and maximum radius of the particles in the disc. The labels give the radius in the middle of each ring. \emph{\bf{The top panels}} show the evolution of the eccentricity (left) and inclination (right) of four representative rings in the $N$-body simulation as dashed lines. \emph{\bf{The bottom panels}} show the same quantities for the SPH simulation. The dotted lines represent the theoretically expected eccentricity and inclination for each ring, using Eq. \ref{eq:acc_grav_tilt} for the $N$-body and Eq. \ref{eq:at_sph_tilt} for the SPH simulations with the average parameters of each ring. Once a ring is fully stripped, we no longer plot the theoretical values. The vertical dashed lines in the bottom panels give the time scale for eccentricity changes in the ring in the corresponding colour according to Eq. \ref{eq:taue_tilt}. \label{fig:V20N6i45_tilt}}
\end{figure*}

We discuss simulation V20N6i45 in detail because it has the highest momentum flux and therefore the change in inclination within the time frame of our simulations is more prominent than in the other models. At $t = 0$, we  binned the disc particles in ten concentric rings. We followed the particles in each ring over the time span of the simulation and averaged the eccentricities and inclinations of the particles in that ring. In Fig. \ref{fig:V20N6i45_tilt} we show the evolution of the eccentricity and inclination of four representative rings within the disc, i.e. the inner and outer ring and two rings that are equally spaced  between them, for both the $N$-body and SPH simulation. In this simulation, disc material at radii $\gtrsim 50$ AU is stripped from the disc by ram pressure. Figure \ref{fig:V20N6i45_tilt} shows that in the $N$-body simulation the rings at 95 and 67 AU are indeed completely stripped after 110 and 390 years, respectively. However, in the SPH simulation a small fraction of the disc material at large radii remains part of the disc and migrates to smaller radii. Less than 1\,\% of material that was originally in the outer rings survives and remains in the disc.

\subsubsection{$N$-body simulation}\label{sec:res_case_nbody_tilt} 
In the $N$-body simulation all rings behave independently, as expected. The larger the distance of the ring to the central star, the faster it changes its eccentricity (cf. Eq. \ref{eq:dedt_tilt}, i.e. both $a$ and $A(r,t)$ increase with $r$). Initially, the eccentricity increases, which is followed by a change in the inclination. This is consistent with the mechanism discussed in Sec. \ref{sec:stark_tilt}, i.e. the inclination changes as a result of the non-zero eccentricity. Although the solutions to the Stark problem presented in Sect. \ref{sec:disctheory_tilt} are simplified and the particles within each ring do not have the same cross section, the initial evolution of the rings is still well described by Eqs. \ref{eq:ecc_tilt} and \ref{eq:acc_grav_tilt}. As expected, in the long term these equations no longer provide an accurate description. We note that the amplitude of the oscillation decreases. This is caused in part by the binning of the particles, which averages out the extreme values. In addition, individual particles reach a maximum eccentricity that is smaller than theoretically expected, i.e. $e_{\rm max} = \sin{i_0}$ (Eq. \ref{eq:ecc_tilt}), on longer time scales. According to the solutions to the classical Stark problem, the minimum and maximum eccentricity do not depend on the semi-major axis of the particle's orbit and the angular momentum component of the particles along the flow direction, which is $L_x$ in our simulations, does not change. However, the stripping of the outer rings also reduces $L_x$ (see Sect. \ref{sec:res_am_tilt}), which might explain the lower maximum eccentricity obtained. This argument is supported by the fact that in simulation V10N5i45, where practically no stripping occurs, particles continue to reach the maximum eccentricity until the end of the simulation.

Apart from the discrepancies described above, the initial evolution of the inclination is fairly well described by Eq. \ref{eq:inc_tilt}.  At later times, the inclination also deviates from the theoretically expected values. We conclude that the solutions to the Stark problem of Sect. \ref{sec:disctheory_tilt} provide insight into the relevant mechanisms for the tilt process of a protoplanetary disc at early times, while on the long term the description is no longer accurate, due to the effects of mass accretion.

\subsubsection{SPH simulation}\label{sec:res_case_sph_tilt} 

\begin{figure*}[!ht]
\centering
\includegraphics[width=\textwidth]{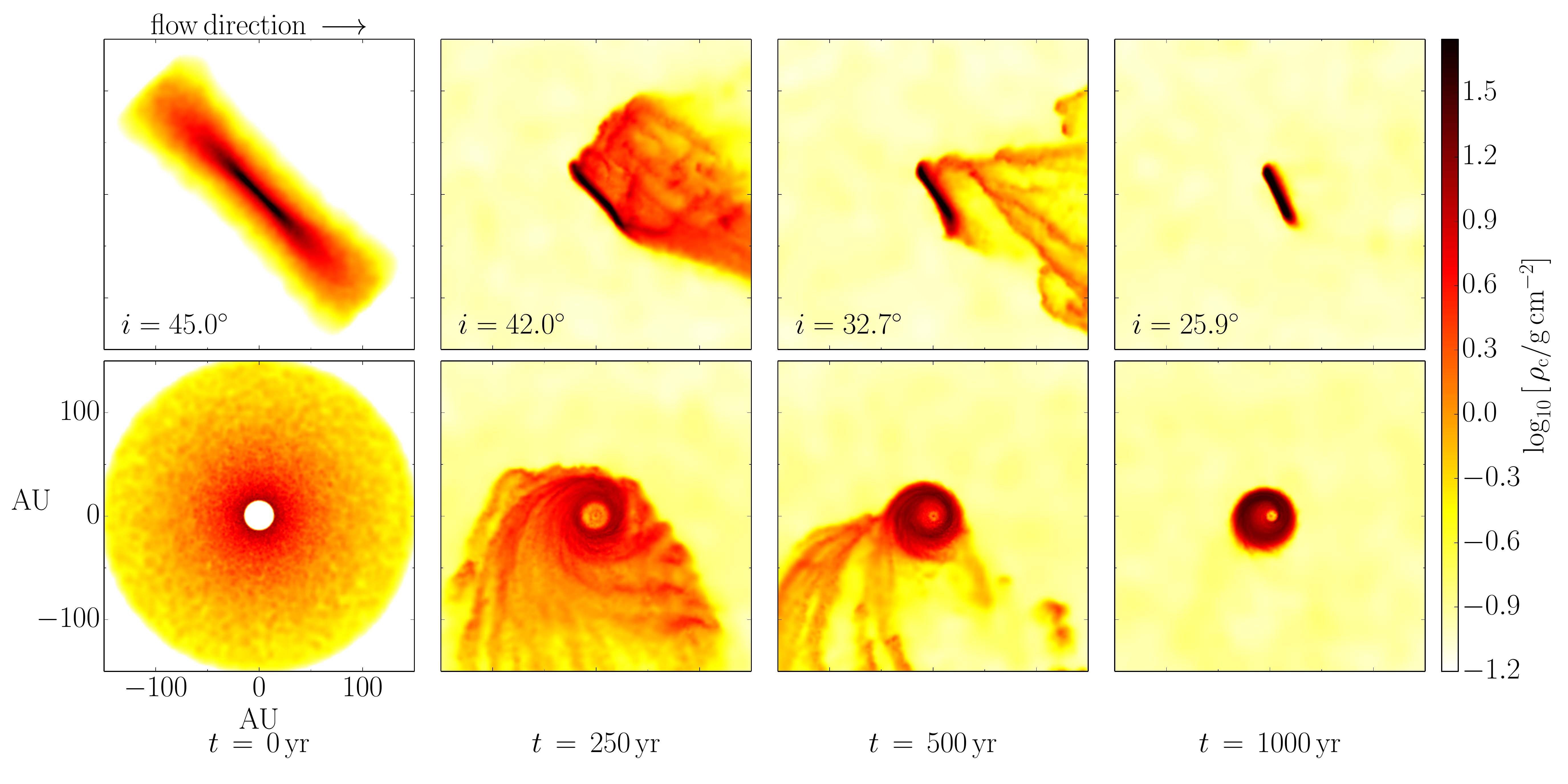}
 \caption{Edge-on \emph{\bf (top row)} and face-on \emph{\bf (bottom row)} snapshots from the V20N6i45 simulations. We note that the face-on view is in the plane of the disc. The
column density, $\rho_c$, is shown in logarithmic scale and is integrated along the full computational domain. The spatial scale is indicated at the bottom left and the flow direction at the top left.}\label{fig:snapshotsv20n6i45_tilt}
\end{figure*}

Upon inspection of the results of the SPH simulation, we see that initially the rings follow a qualitatively similar behaviour to the $N$-body runs and the theory of Sect. \ref{sec:disctheory_tilt}. The rings react independently to the ISM flow during the first 800 years. At later times, in contrast to the $N$-body simulation, the eccentricities and inclinations are equalised and the disc behaves collectively. We see this in all the simulations, but the time scale on which the disc starts behaving as a whole differs for each simulation. The higher the momentum flux in the simulation, the shorter the time scale on which the rings start to evolve synchronously. 
Another notable difference between the $N$-body and SPH simulations is that the amplitude of the oscillation is strongly damped in the latter. This difference cannot be attributed to the averaging of the eccentricity of particle orbits within a ring, but it is due to gas dynamical and viscous interactions, which limit the eccentricities that the orbits of the particles attain. As a result of dissipation within the disc, the eccentricity cannot increase to the maximum value given by the classical Stark problem. Eq. \ref{eq:dedt_tilt} should contain a dissipation term if it were to describe the evolution of the gaseous disc more accurately. 
The dissipation in the disc eventually circularises the orbits. The maximum eccentricity obtained by a ring depends on the time scale on which the eccentricity increases: at large radii the eccentricity
increases faster and therefore has reached a higher value before dissipation causes the eccentricity to decrease. This is visible in the bottom left panel of Fig. \ref{fig:V20N6i45_tilt}, where the outer rings reach a higher eccentricity than the inner rings.

The initial change in inclination of the rings depends on their radii, in accordance with the evolution of the eccentricity. The increasing average inclination of the two outer rings, starting around 150 years, is an artefact caused by disc material that is in the process of being stripped but is still associated with the disc by the algorithm that we use to distinguish the disc from the ISM flow. At around 500 years, when this material is completely removed from the disc, the inclination of these rings decreases rapidly as the small fraction of gas they retain now determines their average inclination. The initially independent behaviour of the outer rings is also visible in snapshots from the simulation. In Fig. \ref{fig:snapshotsv20n6i45_tilt} we  show snapshots of the column density in the V20N6i45 simulation at four consecutive times: $t=0$,  250, 500, and 1000 years. The top row shows the edge-on view where the ISM is flowing in from the left and the bottom row shows the face-on view   looking along (and in the direction of) the angular momentum vector of the disc (as in Fig. \ref{fig:stark_tilt}). When the flow hits the disc, the outer edge of the disc is stripped by ram pressure. At $t=250$ and 500 years, the outer edge  experiences a larger change in its inclination than the rest of the disc. At $t=1000$ years the disc behaves as a whole, consistent with Fig. \ref{fig:V20N6i45_tilt}.
In the face-on view at $t=250$ years the increasing column density at small radii shows that the disc is contracting rapidly. This is caused by  the accretion of ISM with no azimuthal angular momentum and  by the change in the inclination, as discussed in Sect. \ref{sec:disctheory_tilt}. Furthermore, the face-on view at $t=1000$ years shows that the disc is clearly lopsided. The snapshots and orbital elements of the gaseous disc indicate that the disc qualitatively behaves in the way illustrated in Fig. \ref{fig:stark_tilt}.

\subsection{Disc evolution for different flow conditions}\label{sec:res_all_tilt}

\begin{figure*}[!t]
\centering
\includegraphics[width=\textwidth]{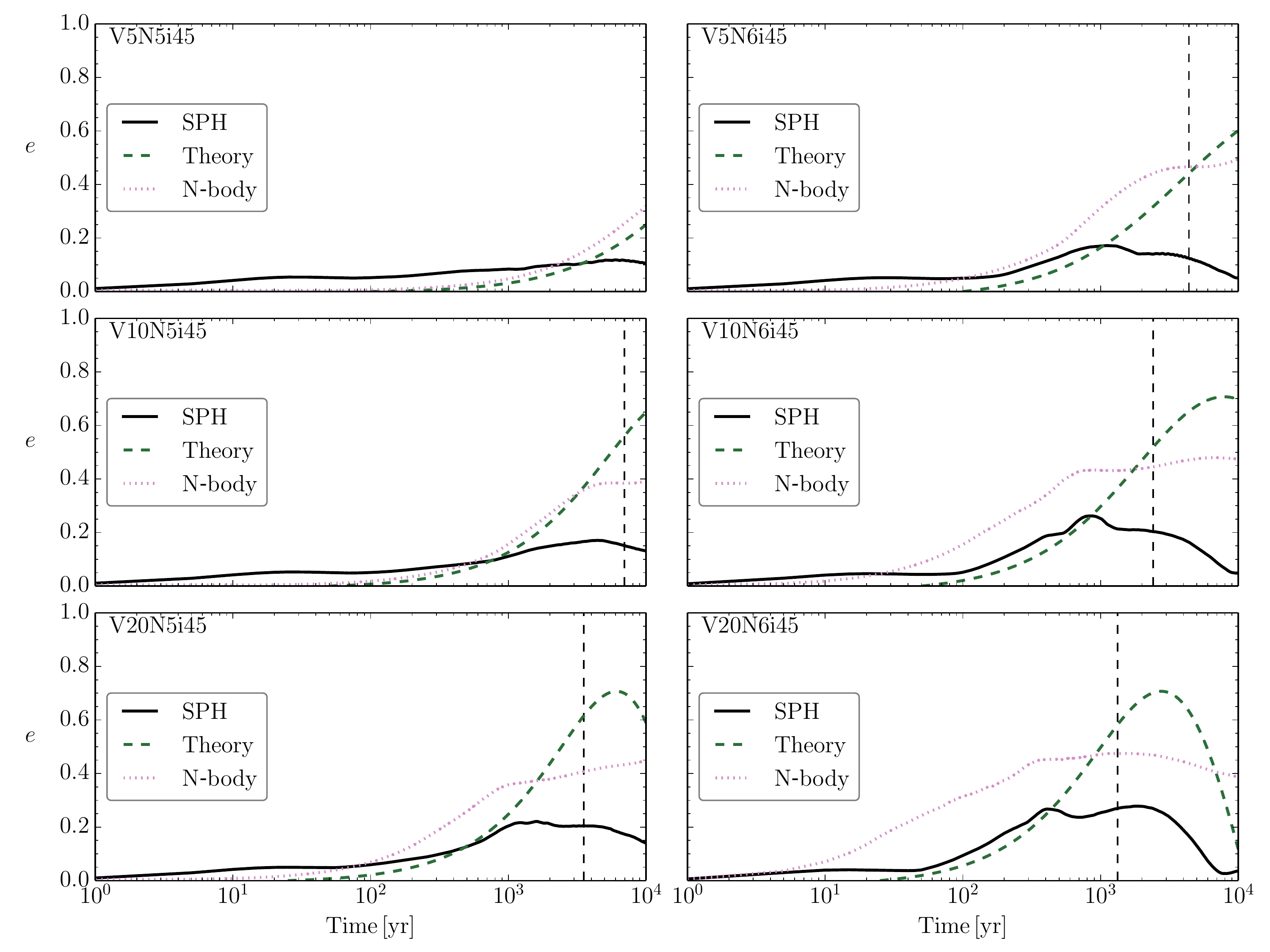}
 \caption{ Mean eccentricity of the particles in the disc for the SPH simulation (black), in the $N$-body simulation (purple), and as expected from the theory presented in Sect. \ref{sec:disctheory_tilt} (green, using Eqs. \ref{eq:ecc_tilt} and \ref{eq:acc_sph_tilt}). The vertical dashed line indicates $\tau_e$ as calculated from Eq. \ref{eq:taue_tilt} for half the truncation radius (see Sect. \ref{sec:res_all_tilt}). The left column shows the simulations with a density of $1.9 \times 10^{-18}$ g/cm$^3$ and the right column the simulations with a density of $1.9 \times 10^{-17}$ g/cm$^3$.}\label{fig:ecc_tilt}
\end{figure*}

\begin{figure*}[!t]
\centering
\includegraphics[width=\textwidth]{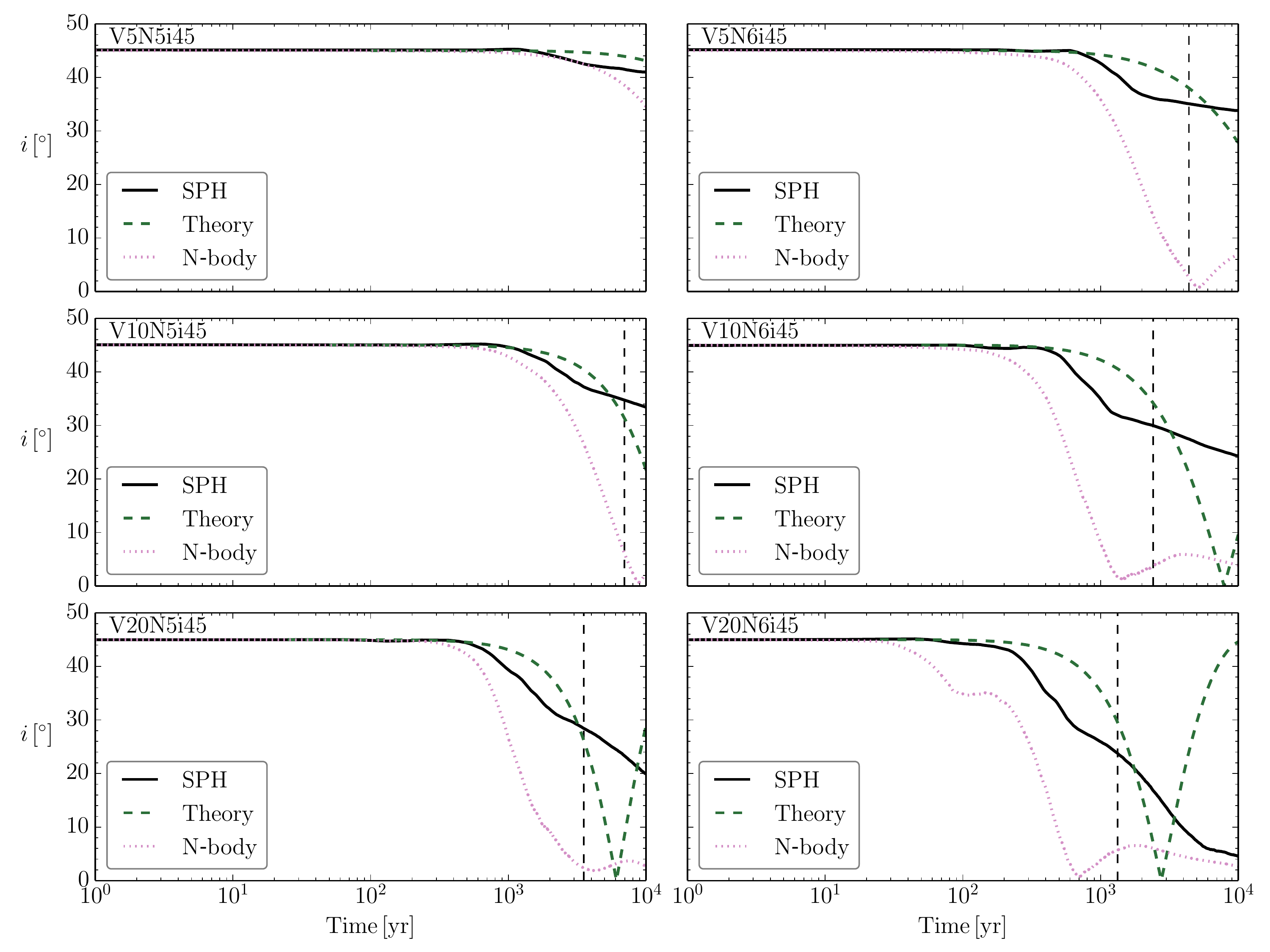}
 \caption{Similar to Fig. \ref{fig:ecc_tilt}, but for the average inclination of the particles in the disc, using the same colour-coding and plotting the same time scales. The theoretically expected inclination is derived from the eccentricity in Fig. \ref{fig:ecc_tilt} using Eq. \ref{eq:inc_tilt}.}\label{fig:inc_tilt}
\end{figure*}

In Figs. \ref{fig:ecc_tilt} and \ref{fig:inc_tilt} we show the evolution of the eccentricity and inclination of the disc for all $N$-body and SPH simulations. We have characterised the disc in each simulation by taking the average of the eccentricity and inclination of all the particles in the disc at each moment in time. We show in Sect. \ref{sec:res_case_tilt} that for the SPH simulations the mean eccentricity and inclination is a good indicator of the evolution of the entire disc once all the rings evolve synchronously. The $N$-body simulations are shown for comparison to the SPH simulations, although the mean eccentricity and inclination do not provide a reliable measure of the long-term behaviour of the individual rings in these simulations (see Sect. \ref{sec:res_case_nbody_tilt}). The vertical line shows the time scale for eccentricity changes (Eq. \ref{eq:taue_tilt}) for the characteristic radius of the disc, which we have chosen to be half the truncation radius of each disc\footnote{The truncation radius is defined as the largest radius in the disc that is not affected by ram pressure stripping. We use Eq. 2 from paper II, taking the ISM velocity component perpendicular to the disc, to calculate this radius and take the minimum of this radius and the initial disc radius.\\The ISM impacts the disc after the start of the SPH simulations, and we correct the time scale for this, which corresponds to an increase of  3\,\% at most.}. The time scale for simulation V5N5i45 is $2.8 \times 10^4$ years, which is longer than the duration of the simulation. Figs. \ref{fig:ecc_tilt} and \ref{fig:inc_tilt} show that the time scale of Eq. \ref{eq:taue_tilt} provides a reasonable indication for the initial change in the average eccentricity and inclination in the SPH simulations. The green dashed curves are based on Eqs. \ref{eq:ecc_tilt} and \ref{eq:acc_sph_tilt} for the SPH simulations, again using half of the truncation radius. This is expected to describe the short-term behaviour of the eccentricity and inclination of the disc in the SPH simulations. 

In the SPH simulations, the gas dynamical and viscous processes in the disc slow down the evolution of the eccentricity and inclination compared the purely gravitational $N$-body simulations. The SPH simulations show an increasing trend in the maximum eccentricity of the disc with increasing momentum flux. As discussed in Sect. \ref{sec:res_case_sph_tilt}, the maximum eccentricity depends on the time scale on which the eccentricity changes, which decreases with increasing momentum flux, and the dissipation time scale within the disc. Assuming that the dissipation time scale is similar in all simulations since the initial disc is the same, it is to be expected that at a higher momentum flux the disc obtains a higher eccentricity before dissipation within the disc imposes the circularisation of the orbits. The maximum eccentricity is reached around the point where the disc starts behaving collectively. After the eccentricity has reached its maximum value, the inclination follows a different trend than theoretically predicted and decreases more slowly. In SPH simulation V10N6i45 the mean eccentricity is 0.05 at the end of the simulation, when the inclination of the disc is still 24\,$^{\circ}$ (see Fig. \ref{fig:inc_tilt}). As long as the eccentricity remains non-zero there will be a net torque on the disc, as is evident from the declining trend of the inclination in Fig. \ref{fig:inc_tilt}. On long time scales, i.e. when $\tau_e > \tau_{\dot{m}}$, the effects of mass loading and contraction have increased the surface density profile and the resulting acceleration is small, making it increasingly hard for the ISM flow to increase the eccentricity again. The dissipation within the disc and the acceleration exerted by the ISM flow probably obtain an equilibrium in which the eccentricity of the disc asymptotically approaches zero. In simulation V20N6i45, the disc has almost reached perpendicular alignment and the eccentricity is 0.05. Therefore, the net torque on the disc is small. As the net acceleration decreases, it is difficult for the gaseous disc to reach completely perpendicular alignment to the flow.

\begin{figure*}[!t]
\centering
\includegraphics[width=\textwidth]{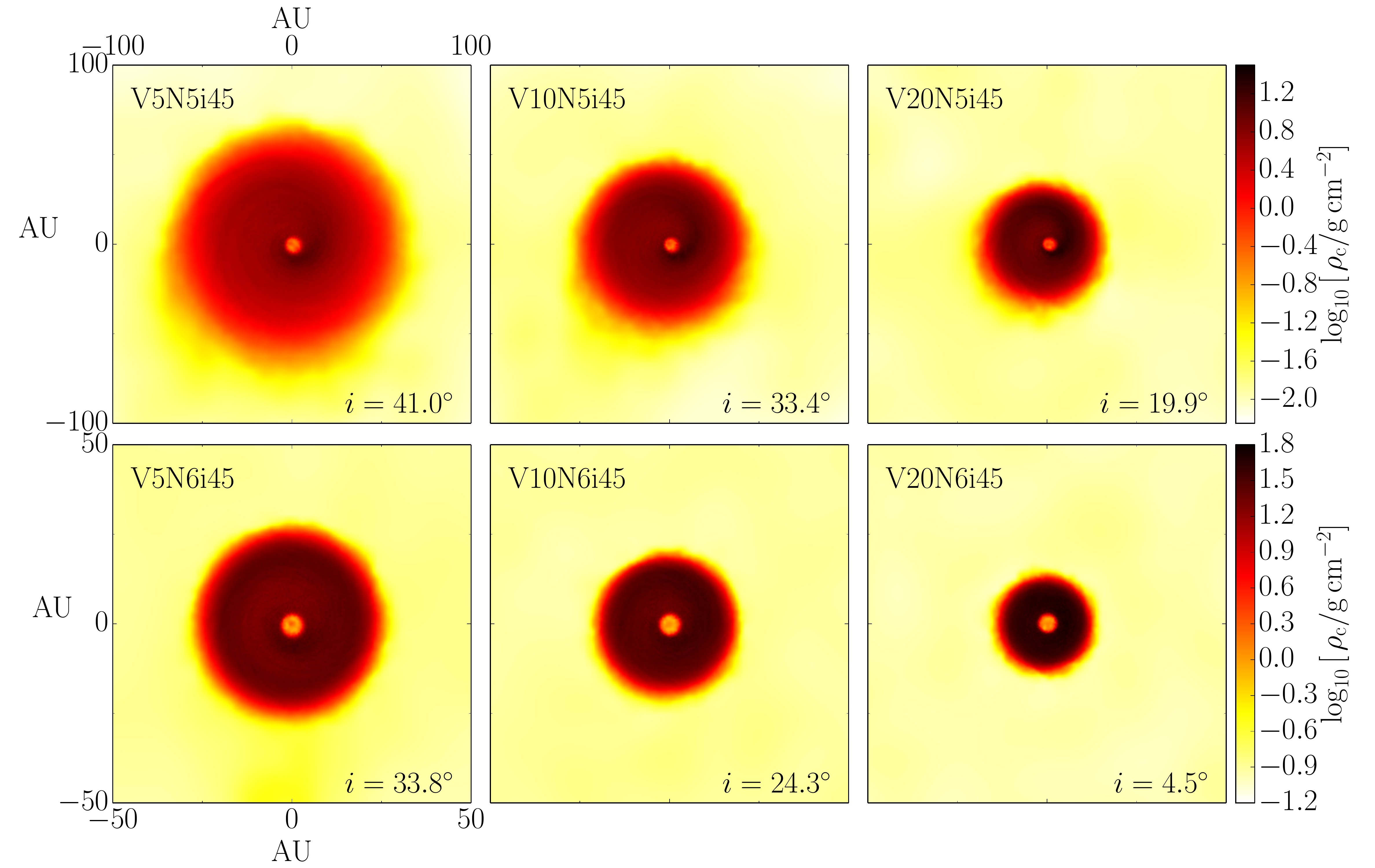}
 \caption{Final snapshot of each SPH simulation at 10 000 years, shown in the plane of the disc. \emph{\bf{The top row}} shows the column density of the simulations with a density of $1.9 \times 10^{-18}$ g/cm$^3$ and velocities of 5, 10, and 20 km/s from left to right. \emph{\bf{The bottom row}} shows the simulations with the same velocities and the higher density of $1.9 \times 10^{-17}$ g/cm$^3$. We note that the spatial and colour scales are different for each row to provide a better contrast for the spirals in the discs.}\label{fig:snapshots_tilt}
\end{figure*}

In Fig. \ref{fig:snapshots_tilt} we present the final face-on snapshot of the discs in our SPH simulations at 10 000 years. We show the column density in the plane of the disc, i.e. the same viewpoint used in Fig. \ref{fig:snapshotsv20n6i45_tilt}. As can also be seen in Fig. \ref{fig:ecc_tilt}, the discs in the simulations with the highest density (bottom row) are almost axially symmetric at the end of the simulation. On the other hand, the discs in the simulations with the low density, in particular V20N5i45, are lopsided and show asymmetries. A one-armed spiral structure is clearly visible in these low-density simulations. The spiral structure is also present in simulations V5N6i45 and V10N6i45, but the density contrast is less sharp at the end of the simulation. All the modelled discs in our simulations show a one-armed spiral during their simulation. The time scale on which the spiral structure forms is roughly the eccentricity growth time scale (Eq. \ref{eq:taue_tilt}), which is consistent with the eccentricity being dissipated in the spiral arm structure.

\subsection{Orientation of the angular momentum vectors}\label{sec:res_am_tilt}

\begin{figure}[t]
\centering
    \includegraphics[width=0.49\textwidth]{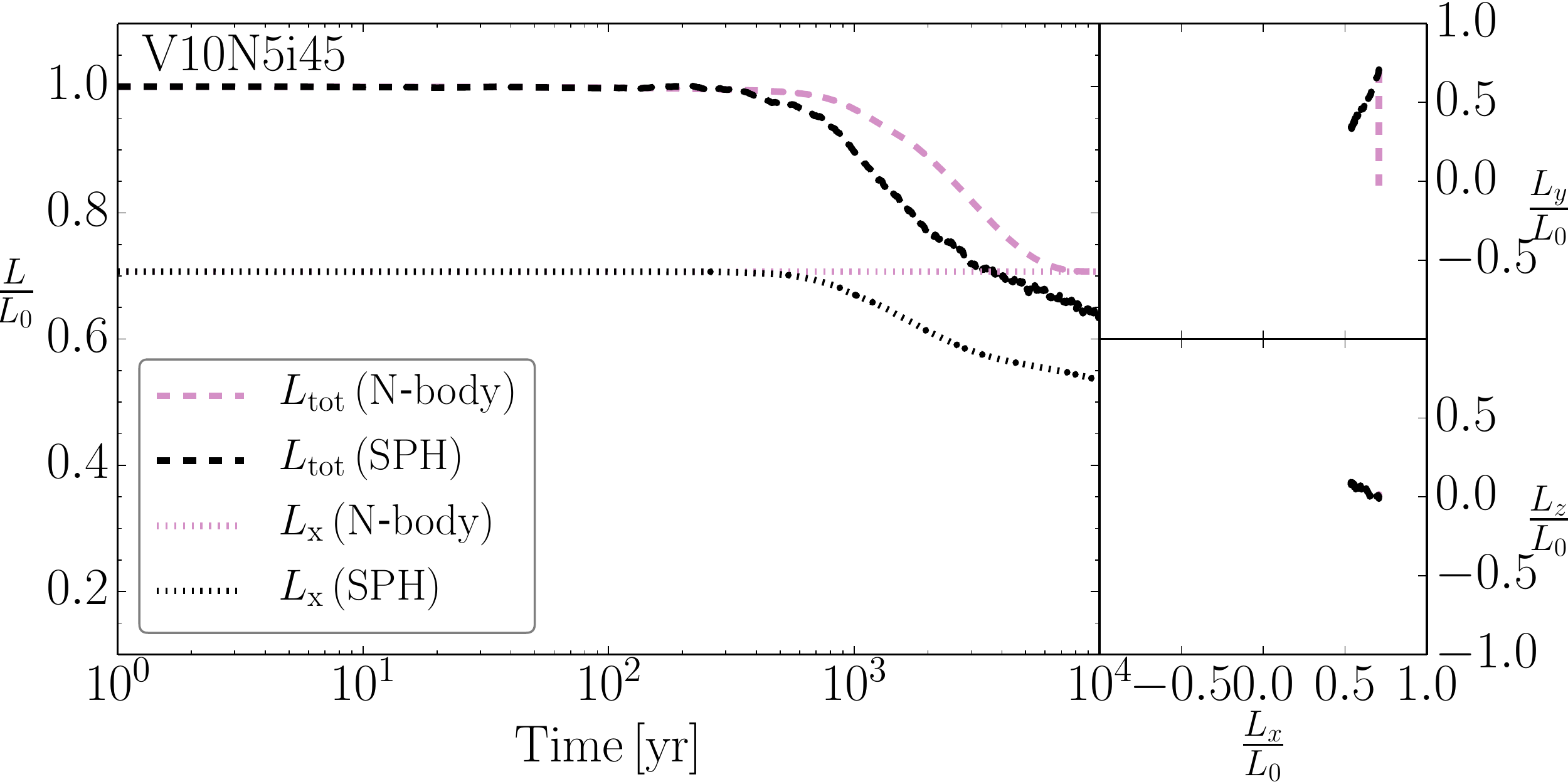}
\caption{\emph{\bf Left frame:} Total angular momentum of the disc, $L_{\rm tot}$, in dashed lines and the component along the flow direction, $L_x$, in dotted lines as a function of time for both the $N$-body (purple) and SPH (black) simulation V10N5i45. \emph{\bf Right square frames:}  Evolution of the two remaining angular momentum components as a function of $L_x$ for the $N$-body and SPH simulations using the same colour coding as in the left frame. All angular momentum vectors are expressed in terms of the initial total angular momentum of the disc.}\label{fig:am_tilt}
\end{figure}

From the solutions to the Stark problem presented in Sect. \ref{sec:stark_tilt}, the total angular momentum of the disc is expected to decrease as it tilts to an inclination of $0^{\circ}$. The change in orientation of the disc therefore causes the disc to contract, in addition to the contraction caused by the accretion of ISM. When no accretion or stripping of disc material occurs, the component of the angular momentum parallel to the velocity vector of the flow is expected to remain constant. Furthermore, there should be no precession of the angular momentum vector. 

To test these predictions, we show the evolution of the angular momentum vectors of simulation V10N5i45 in Fig. \ref{fig:am_tilt}. We  express all angular momentum components in terms of the initial value of the total angular momentum of the disc. The large frame on the left shows the evolution of the total angular momentum of the disc, $L_{\rm tot}$, and the angular momentum component along the flow direction, $L_{x}$, as a function of time for both the SPH and $N$-body simulation. The two smaller frames on the right show the evolution of the other two angular momentum components, $L_{y}$ and $L_{z}$, as a function of $L_{x}$. 

In the $N$-body simulation, where no particles become unbound, $L_x$ remains constant, while the total angular momentum decreases. When the disc is practically aligned perpendicular to the flow, $L_x = L_{\rm tot}$ and both $L_y = L_z = 0$ and all components remain constant during the rest of the simulation. The evolution of $L_y$ and $L_z$ as a function of $L_x$ show that   no precession is occurring.

In the SPH simulation, the situation is complicated by the continuous loss of angular momentum. Gas from the disc is accreted onto the star and disc material is continuously stripped from the outer edge of the disc. Previous work suggests that this stripping is numerical rather than physical (see Sect. 5.2 of paper II). Nonetheless, even keeping this continuous loss of angular momentum in mind, Fig. \ref{fig:am_tilt} illustrates that in the SPH simulation, the same process occurs. The angular momentum vectors do not precess and although $L_{x}$ does not remain constant, its change is caused by continuous stripping of disc material rather than by the tilt process.

\section{Discussion}\label{sec:discussion_tilt}

\subsection{Model uncertainties}

Although we use simplified solutions to the Stark problem, we have shown that essentially the same physical mechanism changes the orientation of a gaseous disc that is subjected to an ISM flow. Using a more detailed description for the semi-major axis $a(t)$ and acceleration $A(t)$ would not provide a proper quantitative description for the evolution of a gaseous disc because the dissipation of the eccentricity within the disc is an unknown function of density, temperature, and viscosity.

The viscosity in our models is not constrained by first principles, but is set by a numerical free parameter and therefore does not provide a physical understanding of the dissipation. In reality, the dissipation in the disc may be slower than in our simulations, in which case the obtained eccentricity will be higher.

The $N$-body models presented in this work do not include self-gravity between the test particles. This is a deliberate choice. The simplified $N$-body model illustrates the basic physical mechanism that changes the orientation of the disc. We  also performed $N$-body simulations taking the gravity between the individual particles into account, but they do not provide additional insight into the process that occurs in the SPH simulations, which is determined by the additional effects of pressure and viscosity.

The tilt process requires that the obits in the disc are (close to) Keplerian. If this it not the case, for example in massive discs, then the eccentric orbits are not closed and they will precess. The resulting rotation of the argument of pericentre affects the direction of the torque. This may prevent the tilt process from occurring, for example, in galactic discs.

\subsection{Observed orientations of protoplanetary discs}\label{sec:dis_orientation_tilt}

\begin{figure}[!t]
\centering
\includegraphics[width=0.49\textwidth]{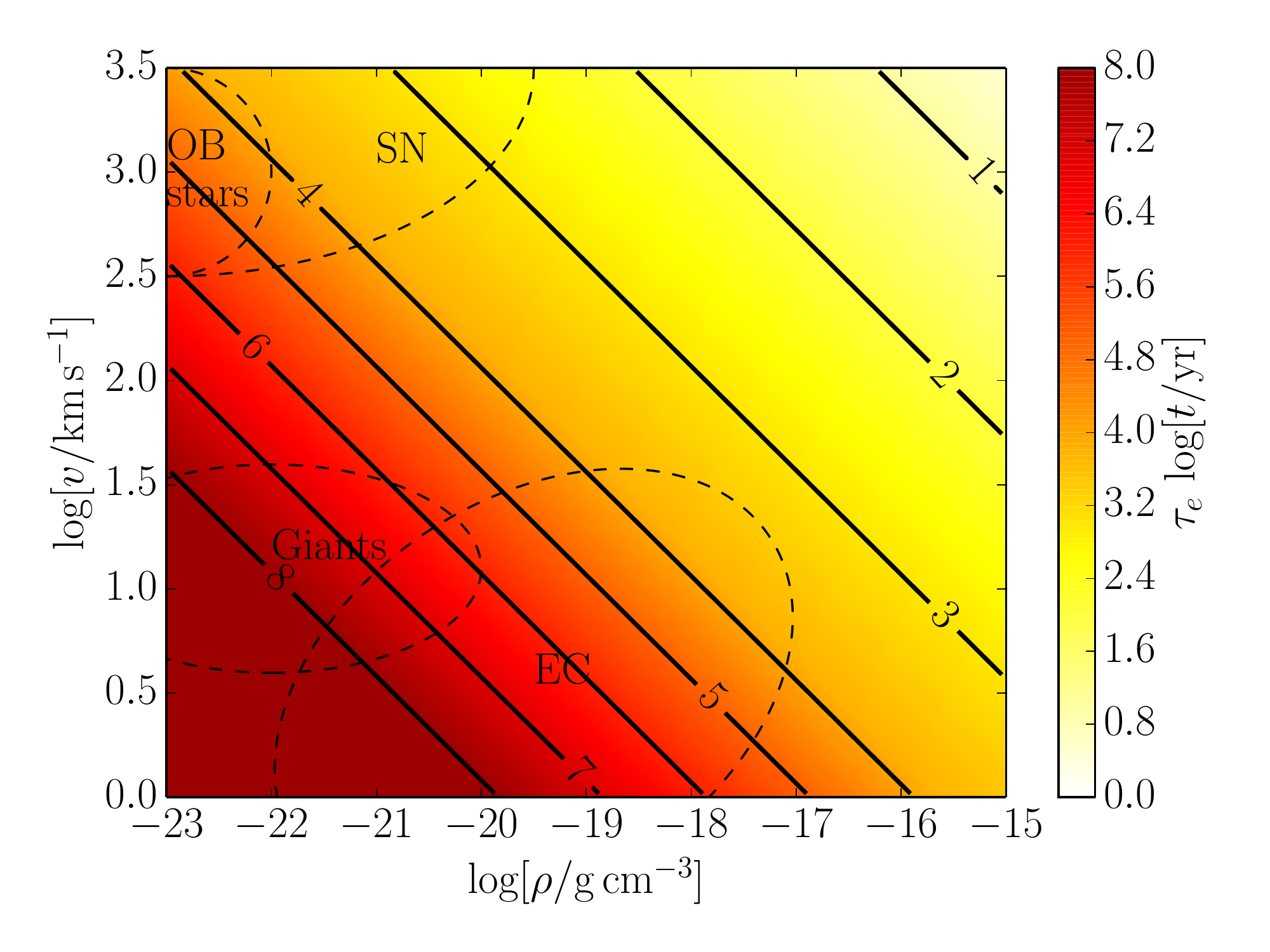}
\caption{Time scale on which the inclination is expected to change and spiral arms are expected to form for the disc parameters, and corresponding typical radii (see Sect. \ref{sec:res_all_tilt}), in our simulations as a function of their ambient density and relative velocity (Eq. \ref{eq:taue_tilt}). The spacing of the isochrones depends on the influence of ram pressure stripping. The circled regions provide an order of magnitude estimate for different environments: winds from massive stars and giants \citep[][Van der Helm et al., in prep.]{owocki13}, supernova ejecta \citep[SN,][]{ouellette07}  and embedded open and young massive clusters \citep[EC,][]{lada03, portegies_zwart10}. \label{fig:taue_tilt}}%
\end{figure}

Figure \ref{fig:taue_tilt} shows  the time scale on which a disc with the same parameters as in our simulations is expected to change its inclination substantially as a function of the ambient density and relative velocity. In addition to the density and velocity, this time scale also depends on the surface density profile of the disc, its initial inclination with respect to the velocity vector, and mass of the host star. Our simulations show that this time scale is an upper limit because the disc generally reacts on a shorter time scale (cf. Fig. \ref{fig:inc_tilt}). The time scale can differ by orders of magnitude between very rapid ($\lesssim 10^4$ years) and slow ($10^6 - 10^7$ years) compared to the typical lifetime of a protoplanetary disc. Typical velocities (1-3 km/s) and average densities ($\sim 10^{-19}$ g/cm$^3$) observed in star-forming regions (see e.g. \citealt{lada03}) are probably not sufficient for a substantial change in inclination. Continuously moving in a straight line through a homogeneous density of $10^{-19}$ g/cm$^3$ with a velocity of 3 km/s will effectively change the inclination after $10^6$ years. Although the densities and velocities in the dense cores of such star-forming regions are sufficient, the stellar orbits change the velocity vector on shorter time scales, i.e. less than the typical cluster crossing time. Therefore, if a population of protoplanetary discs in a star-forming region is observed to have a preferred orientation with respect to their velocity vector, this may indicate that their velocities and the ambient density that the discs have moved through in the past were high. This can provide constraints on the initial conditions of the star-forming region, in particular in the outskirts of dense star-forming regions where the frequency of close encounters that change the orientation of the velocity vector is low. However, in the outskirts the ambient gas density is also lower and the change in inclination may be small. In the core of the cluster, the imprint of the tilt process may be washed out by the frequent orbital changes of the stars. Even if the frequent change of the velocity vector in the core of an embedded cluster had a small net effect on the inclination of the disc, the cumulative effect of the small inclination changes could still lead to a substantial contraction of the disc.

The tilt process is expected to have little effect in small and sparse star-forming regions where the densities and velocities are low. Because a higher mass density implies a higher velocity dispersion, a correlation is expected between the mass density of star-forming regions and the orientation of the protoplanetary discs they host. This is in contrast with other scenarios that explain the observed spin-orbit misalignment as these generally do not prefer a certain orientation (e.g. dynamical encounters) or do not depend on the environment (e.g. magnetic interactions between the star and the disc). The relevance of the tilt process with respect to other scenarios can be tested with  observations and with simulations. Simulations of stellar orbits in star-forming regions can provide insight into whether the expected distribution of disc orientations is distinguishable from a random distribution. 

The presence of winds from massive and giant stars and supernovae ejecta can also affect the orientation of nearby discs. The winds of giants have typical velocities of the order of 10 km/s, while for massive stars and supernova ejecta this is two to three orders of magnitude higher, which increases the relative velocity and makes the process more effective (see Fig. \ref{fig:taue_tilt}). The temperature of these ejecta is much higher than assumed in this work. As a result, the disc may not accrete the hot gaseous ejecta as efficiently as it accretes the cold gas. \citet{ouellette07} performed {2D} simulations of the interaction of a protoplanetary disc with supernova ejecta and found that 99\,\% of the hot gaseous material is deflected around the disc \citep[contrary to the cold dust of which more than 90\,\% is accreted,][]{ouellette09, ouellette10}. However, the tilt process depends on the force exerted by the inflowing material and not on the accreted material. The net effect on the tilt process may be similar, but to verify this requires additional simulations that are beyond our computational limitations (see Sect. \ref{sec:dis_solarsystem_tilt}).

Observations show that most hot Jupiters, i.e. Jupiter-mass planets with a period $\lesssim 10$ days, are misaligned \citep[see e.g.][]{triaud10}, but the spin-orbit misalignment is not restricted to hot-Jupiter systems \citep{huber13}. The formation of hot Jupiters is generally associated with dynamical interactions between planetary and stellar companions. The observations that hot Jupiters around main sequence stars occur more frequently in open clusters than in the field \citep{brucalassi16, brucalassi17} is therefore related to dynamical interactions within the cluster. The contraction of protoplanetary discs caused by the tilt process and accretion of ISM would lead to more compact and possibly misaligned planetary systems in dense environments compared to planetary systems in sparse regions and the field \citep{wijnen17}. A study of the occurrence of spin-orbit misalignment as a function of stellar environment may provide additional constraints on the relevance of the tilt process, in particular for stars with multiple co-planar planets. Based on the time scale estimate in Fig.~\ref{fig:taue_tilt} and on our simulations the tilt process will probably not cause very large obliquities, which are more likely the result of dynamical interactions.

\subsection{Asymmetries and spiral arms in protoplanetary discs}\label{sec:dis_asymmetries_tilt}

Our simulations show that asymmetries and spiral structures observed in protoplanetary discs, which are usually associated with planet formation, could be caused by the movement of a protoplanetary disc through an ambient medium. In turn, these structures could cause or aid planet formation. The spiral structures and asymmetries form even when the inclination of the protoplanetary disc does not change substantially, as in simulation V5N5i45. The spiral arms in simulations V5N5i45 and V10N5i45 are similar, in both  shape and size, to the spiral-like structure observed by \citet{van_der_marel16} in the HD135344B transition disc (see their Fig. 1). The time scale given by Eq. \ref{eq:taue_tilt} gives us an indication of the extent to which the tilt process can cause the formation of spirals arms and asymmetries in the disc in nature. The tilt process can only be plausibly considered to be a trigger for planet formation if it occurs on time scales $\lesssim 10^6$ years, as planets are expected to start forming around protoplanetary disc lifetimes of $10^6$ years \citep[e.g.][]{williams11}. As discussed in Sect. \ref{sec:dis_orientation_tilt}, for typical velocities in star-forming regions the protoplanetary disc will have to move through an ambient density of $\gtrsim 10^{-19}$ g/cm$^3$ continuously for $10^6$ years, which can only occur in the cores of embedded clusters. Therefore, for spiral arms to form in this process, the protoplanetary discs should pass through a high-density region and/or the relative velocity should be high. This passage can be relatively brief with a duration of $\lesssim 10^4$ years. This may happen in the vicinity of winds from massive stars and supernova ejecta,  although that introduces stochasticity making it unlikely as a general process. A more generic situation would arise when the star and protoplanetary disc leave their natal core. Whether this is a plausible scenario needs to be investigated with future studies, for example by integrating the orbits of stars and their protoplanetary disc over their ambient density and relative velocity in simulations of star-forming regions \citep[e.g.][]{bonnell03}. 

Our study also suggests that, similar to the orientation, the occurrence of lopsided discs may scale with the gas density and velocity dispersion of star-forming regions. Observational studies on the occurrence of lopsided protoplanetary discs can therefore also shed light on the relevance of this process. Simultaneously, such observations could provide constraints on the time scale of dissipation of the eccentricity in the disc. However, these studies may be difficult because our work suggests that in the long term, $t \gtrsim \tau_e$, the eccentricity of the disc may not be much larger than $\sim$ 0.1.

\subsection{The solar system}\label{sec:dis_solarsystem_tilt}

The observed misalignment of $7^{\circ}$ between the equatorial plane of the Sun and the orbital plane of its planets can, in principle, be explained by the tilt process presented here. The observation of short-lived radioisotope tracers in meteorites \citep[see e.g.][]{tachibana03} indicates that during its formation the solar system may have experienced an interaction with supernova ejecta. Unfortunately, we cannot perform a fully consistent hydrodynamical simulation with our SPH model because the relatively low density and high velocity of the supernova ejecta would cause severe numerical artefacts with our current code. However, we can investigate whether this interaction can also explain the observed misalignment by performing a simulation with our simplified $N$-body model, assuming a 1 $\Msun$ star and otherwise the same disc parameters as in our other $N$-body models. We use the time-dependent supernova density, $\rho \propto \left((t + t_{\rm trav})/t_{\rm trav}\right)^{-3}$, and velocity, $v \propto \left((t + t_{\rm trav})/t_{\rm trav}\right)^{-1}$, from \citet{ouellette07} \citep[see also][] {matzner99}, which scale with the travel time $t_{\rm trav}=d/v_0$ from the supernova to the disc where $v_0$ is the initial velocity of the ejecta, and assume that all momentum is transferred but only 1\,\% of the injected mass is accreted. For a supernova explosion with an energy of $10^{51}$ ergs and ejected mass of $20\,\Msun$ at a distance of 0.3 pc, our simulation gives an average eccentricity of 0.2 and a change in inclination of $9^{\circ}$ when the ejecta have passed after 2000 years, which is the duration of the simulation in \citet{ouellette07}. Within the first 100 years 80\,\% of the change in inclination is obtained, which approximately corresponds to the scaling time scale $t_{\rm trav}$. The reaction of the disc in the $N$-body simulations is generally faster than in the hydrodynamical simulations, but the estimated time scale for supernova ejecta in Fig. \ref{fig:taue_tilt} overlaps with the 2000 years given by the \citealt{ouellette07} model. Furthermore, the radius of the disc is truncated to roughly 40 AU in the $N$-body simulation, which resembles the inferred outer edge of the early solar nebula of 30 AU \citep[see e.g.][]{adams10}. The eccentric orbits that result from interaction with the supernova ejecta may be related to other phenomena observed in the solar system. When planetesimals move on eccentric orbits, an inclination instability may set in (see \citet{madigan16}), which can explain the observed strange orbits of planetesimals in the outer solar system without needing to invoke  `Planet 9' (\citealt{batygin16}, although see \citealt{jilkova15} for an alternative scenario in which these objects could have been captured during a fly-by). Further accurate modelling of the interaction of the Sun's protoplanetary disc with supernova ejecta can provide a more conclusive insight into the possible relation between the enrichment of the solar system and its spin-orbit misalignment. Such studies may also help to put constraints on the nature and parameters of the ejecta that interacted with the solar system.

\section{Conclusions}

In this work we explain why a protoplanetary disc -- and  by extension any circumstellar disc -- changes its orientation and shape when it moves through an ambient medium. The driving mechanism behind the change in inclination is similar to the physical process known as the Stark problem (i.e. eccentricity pumping with conservation of the angular momentum component along the direction of the force) for which theoretical solutions exist. Here we use these solutions and compare them to  a simplified $N$-body model and to self-consistent hydrodynamical simulations. Even if the disc is initially axisymmetric, the force exerted by the flow will make the disc lopsided and cause a net torque on the disc. Consequently, the inclination of the disc changes and the disc will align its angular momentum vector parallel to the relative velocity vector between the disc and the ISM. This process occurs without precession of the disc. The higher the velocity relative to the ISM, the more effectively this process changes the inclination. 

In the restricted theoretical Stark problem, the angular momentum component along the flow direction remains constant, but the net torque decreases the total angular momentum. Therefore, in the process of changing its orientation, the disc also contracts. This contraction is in addition  to the contraction caused by the accretion of ISM with no azimuthal angular momentum with respect to the disc \citep[see][]{wijnen17}. The continuous contraction and accretion of the ISM decreases the efficiency of the tilt process. The higher the surface density profile of the disc, the more difficult it becomes to change its orientation as the force per unit mass decreases. Furthermore, the net torque also decreases with decreasing disc size. 

The increase in eccentricity of orbits in the disc is counteracted by gas dynamical and viscous forces that circularise the orbits. From our simulations, we cannot draw any conclusions about the nature and 
strength of this dissipation, not least because the viscosity in our simulations is numerical. The physical viscosity is approximated because its nature and magnitude are still subject to debate.

In addition to the disc becoming eccentric, a spiral arm forms in the disc maintained by the balance between the eccentricity dissipation and the force exerted by the flow. Spiral arms are usually thought to be invoked by the presence of a planet, but our study shows that this is not a necessary culprit. On the contrary, the emergence of a spiral arm may initiate planet formation. The spiral arm forms even if the inclination of the disc does not change substantially. Protostars and their discs emerge from embedded environments, but it is not clear whether the formation of spiral arms is a generic process in such environments. For the tilting of the disc to occur within about $10^6$ years, relative velocities $\gtrsim 5$ km/s and/or ambient densities $\gtrsim 10^{-19}$ g/cm$^3$ are required. Conditions with similar time scales can be found in young massive clusters and also in regions that have been affected by winds from massive stars or by supernova ejecta. Simulations with our simplified $N$-body model indicate that the interaction of the solar system with supernova ejecta could be responsible for the observed $7 ^{\circ}$ misalignment between the Sun's equatorial plane and the plane of the planetary obits.

\begin{acknowledgements}
We are thankful to Lucie J\'ilkov\'a, Martha Saladino, and Ann-Marie Madigan for valuable discussions. We also thank the anonymous referee. 
This research is funded by the Netherlands Organisation for Scientific Research (NWO) under grant 614.001.202.
\end{acknowledgements}

\bibliographystyle{aa} 
\bibliography{phdbib}

\end{document}